\documentclass[twocolumn]{aastex61}

\usepackage{color}
\usepackage{epsfig}
\usepackage{amsmath}
\usepackage{enumerate}
\usepackage{epstopdf}
\usepackage{verbatim}
\usepackage{natbib}
\usepackage{gensymb}
\usepackage{longtable}
\usepackage{threeparttablex}
\usepackage{tabularx}
\usepackage{booktabs}

\begin{document}

\title{Statistics of coronal dimmings associated with coronal mass ejections. II. Relationship between coronal dimmings and their associated CMEs}

\correspondingauthor{K. Dissauer}
\email{karin.dissauer@uni-graz.at}

\author{K. Dissauer}
\affiliation{Institute of Physics, University of Graz, 8010 Graz, Austria}
\author{A. M. Veronig}
\affiliation{Institute of Physics, University of Graz, 8010 Graz, Austria}
\affiliation{Kanzelh\"ohe Observatory, University of Graz, 9521 Treffen, Austria}
\author{M. Temmer}
\affiliation{Institute of Physics, University of Graz, 8010 Graz, Austria}
\author{T. Podladchikova}
\affiliation{Skolkovo Institute of Science and Technology, 143026 Moscow, Russia}

\begin{abstract}
We present a statistical study of 62 coronal dimming events associated with Earth-directed CMEs during the quasi-quadrature period of STEREO and SDO. This unique setting allows us to study both phenomena in great detail and compare characteristic quantities statistically. Coronal dimmings are observed on-disk by SDO/AIA and HMI, while the CME kinematics during the impulsive acceleration phase is studied close to the limb with STEREO/EUVI and COR, minimizing projection effects.
The dimming area, its total unsigned magnetic flux and its total brightness, reflecting properties of the total dimming region at its final extent, show the highest correlations with the CME mass~($c\sim0.6-0.7$).
Their corresponding time derivatives, describing the dynamics of the dimming evolution, show the strongest correlations with the CME peak velocity \mbox{($c\sim 0.6$)}.
The highest correlation of $c=0.68\pm0.08$ is found with the mean intensity of dimmings, indicating that the lower the CME starts in the corona, the faster it propagates.
No significant correlation between dimming parameters and the CME acceleration was found. However, for events where high-cadence STEREO observations were available, the mean unsigned magnetic field density in the dimming regions tends to be positively correlated with the CME peak acceleration ($c=0.42\pm0.20$). This suggests that stronger magnetic fields result in higher Lorentz forces providing stronger driving force for the CME acceleration. 
Specific coronal dimming parameters correlate with both, CME and flare quantities providing further evidence for the flare-CME feedback relationship. 
For events in which the CME occurs together with a flare, coronal dimmings statistically reflect the properties of both phenomena.
\end{abstract}

\section{Introduction} \label{sec:intro}
Coronal mass ejections (CMEs) and in particular Earth-directed halo CMEs, are the main drivers for severe space weather events affecting the near-Earth environment \citep[e.g.][]{Gosling:1993,Gopalswamy:2010}. However, they allow the least accurate measurements of their properties due to strong projection effects \citep{Burkepile:2004} and especially their early evolution is not observed with traditional white light coronagraphs.
Therefore, the study of associated phenomena low in the corona is essential to obtain additional information on their initiation and early evolution.

The most distinct phenomena associated with CMEs are coronal dimmings, i.e. localized regions of reduced emission in the extreme-ultraviolet (EUV) and soft X-rays (SXR)  low in the corona \citep[e.g.][]{Thompson:1998, Thompson:2000,Hudson:1996,Sterling:1997}. They are assumed to form due to density depletion as a result of the expansion of the CME structure and overlying fields. Several studies confirm this, such as spectroscopic observations of strong material outflows in these regions \citep[e.g.][]{Harra:2001, Tian:2012} and Differential Emission Measure (DEM) studies showing substantial density drops in dimming regions \citep[e.g.][]{Cheng:2012,Vanninathan:2018}. 

During the CME expansion, the overlying field is stretched and partly reconnecting, which is observed as widespread and more shallow dimming regions, so-called secondary dimmings \citep{Attrill:2007, Mandrini:2007}. Core dimmings, on the other hand, mark the footpoints of the erupting flux rope, which is either pre-existing or formed during the eruption via magnetic reconnection. Core dimmings are observed as localized, regions of strongly reduced emission, close to the eruption site in opposite polarity regions \citep{Hudson:1996, Webb:2000}. 

The detailed study of coronal dimmings and their statistical relationship to decisive CME quantities leads to a better understanding of the early evolution of CMEs and may also give insight into the initial configuration of the eruption.
Numerous papers analyzed coronal dimmings in case-studies, investigating their relationship to the CME mass \citep[e.g.][]{Harrison:2000, Harrison:2003, Zhukov:2004, Lopez:2017}, to the morphology and early evolution of the CME \citep[e.g.][]{Attrill:2006, Qiu:2017}, and their timing \citep[e.g.][]{Miklenic:2011}. Only a few studies looked into the coronal dimming/CME relationship statistically \citep{Bewsher:2008,Reinard:2008, Reinard:2009, Mason:2016,Aschwanden:2016,Krista:2017,Aschwanden:2017}.

In this paper, we present a statistical analysis of 62 coronal dimmings and their corresponding CMEs using optimized multi-point observations, where coronal dimmings are observed on-disk by SDO and the associated CME evolution close to the limb by STEREO, minimizing projection effects on the derived CME kinematics.
This allows us for the first time to simultaneously study the time evolution of the CME together with the development of the associated coronal dimming in the low corona.
To this aim, we developed a new method to detect coronal dimmings using logarithmic base-ratio images. This approach allows us to detect and distinguish both types of dimming, core and secondary dimmings \citep{Dissauer:2018a}.

In \cite{Dissauer:2018b}, thereafter paper I, we studied characteristic parameters describing the dimming and its evolution, namely, the dimming area, its total brightness and the total unsigned magnetic flux involved in the dimming regions. These parameters are extracted as time-integrated quantities allowing us to also investigate their dynamics, by calculating the corresponding time derivatives, such as the area growth rate, the brightness change rate and the total magnetic flux rate. In addition, we studied the relation of the dimming parameters with the associated flare, such as the SXR class, the flare fluence and the flare ribbon reconnection fluxes.
This second part of the statistical study focuses on the relationship of coronal dimmings and characteristic CME parameters, such as mass, velocity and acceleration.
\section{Data Set}
\subsection{Event selection}
We focus on Earth-directed CMEs, observed close to the limb with STEREO, and associated with coronal dimmings that occurred on-disk for SDO. Both phenomena are studied using simultaneous multi-point observations minimizing projection effects. The data set consists of 62 dimming events that occurred between 2010 May and 2012 September, i.e.~the time period when SDO and STEREO were located in quasi-quadrature.
Their eruption site lies within $\pm$40\degree~from the central meridian of the Sun and all events are associated with flares, ranging from B to X class. A detailed description how the events were selected can be found in paper I.

\subsection{Data and Data reduction}
To study coronal dimmings we use filtergrams of the Atmospheric Imaging Assembly (AIA; \citealt{Lemen:2012}) and the 720~s line-of-sight (LOS) magnetograms of the Helioseismic and Magnetic Imager (HMI; \citealt{Schou:2012}) on-board the Solar Dynamics Observatory (SDO; \citealt{Pesnell:2012}) and follow the pre-processing as described in paper I.

To measure the kinematics of CMEs and their mass we use data from the STEREO twin spacecraft \citep{Kaiser:2008}. The SECCHI instrument suite \citep{Howard:2008} includes an Extreme Ultraviolet (EUV) Imager \citep[EUVI;][]{Wuelser:2004}, two white light coronagraphs (COR 1 and COR2), and two white light heliospheric imagers (HI1 and HI2). To study the early evolution phase of the CME kinematics, from close to the solar surface up to a distance of 15~R$_{\odot}$, we combine data of EUVI, COR1, and COR2. EUVI observes the solar chromosphere and low corona up to 1.7~R$_{\odot}$. 

The majority of events were studied using images of the 195 \AA~passband with a cadence of 5~minutes. For 20 events, 171 \AA~observations with a time cadence of 75~s were available. This means that only for this subset of the event sample detailed acceleration profiles of the impulsive acceleration phase could be derived.
The inner and outer coronagraphs, COR1 and COR2, have a field-of-view of 1.4 to 4~R$_{\odot}$ and 2.5 to 15~R$_{\odot}$, respectively. The time cadence for COR1 observations was 5 minutes, while COR2 provides observations every 15 minutes.

\section{Methods and Analysis}\label{sec:method}
We study the relationship between coronal dimmings and their associated CMEs.
Section~\ref{sec:dimming} summarizes the dimming analysis that is described in detail in paper I. Section~\ref{sec:cme} describes how the height-time measurements of CMEs are performed and how velocity and acceleration profiles are derived. The calculation of the CME mass is given in Section~\ref{sec:cme_mass}. The methods are illustrated for the dimming/CME events that occurred on 2011 October, 2 (no. 29, impulsive M3.9 flare associated with an EUV wave) and on 2011 February, 13 (no. 6, impulsive M6.6 flare associated with a halo CME and an EUV wave). The time evolution of selected dimming parameters and the CME kinematics, presented in the results section, is also shown for these examples.

\subsection{Dimming analysis}\label{sec:dimming}
Coronal dimmings are extracted from logarithmic base-ratio images using a thresholding algorithm \citep{Dissauer:2018a}.
We analyze different characteristic parameters and study their time evolution in order to describe the dynamics, morphology, magnetic properties and brightness of the total dimming region, including both types of dimming, i.e. core and secondary dimmings.
These quantities are extracted as time-integrated quantities by cumulating newly detected dimming pixels over time. The derivative of each parameter-time profile reflects the parameter's dynamics.

The size of the coronal dimming $A(t_{n})$ is determined by cumulating the area of all dimming pixels that are detected until $t_{n}$, the end time of the dimming evolution. To measure the growth rate of the area $\dot{A}(t_{i})$ at time step $t_{i}<t_{n}$, we calculate the time derivative of the area evolution, respectively. 
We also define the ``magnetic area'' of the dimming $A_{\Phi}(t_{n})$ as the area of all dimming pixels where the magnetic flux density $B$ (measured from SDO/HMI line-of-sight magnetograms) is higher than the noise level of $\pm$10~G. Likewise, the magnetic area growth rate $\dot{A}_{\Phi}(t_{i})$ is extracted as time derivative of $A_{\Phi}$.
The magnetic properties of coronal dimmings are analyzed by the total unsigned $\Phi(t_{n})$, the positive $\Phi_{+}(t_{n})$, and the negative magnetic flux $\Phi_{-}(t_{n})$, their corresponding magnetic flux rates $\dot{\Phi}(t_{i})$ and the mean unsigned magnetic flux density $\bar{B}_{\text{us}}(t_{n})$.

To study the brightness evolution solely as a result of intensity change and independent from variations in the dimming area, $I_{\text{cu,diff}}(t_{i})$ sums the intensity of all dimming pixels within a constant area $A(t_{n})$ at any time step $t_{i}$. Its time derivative $\dot{I}_{\text{cu,diff}}(t_{i})$ reflects the brightness change rate. 
We also extract the mean intensity of the dimming by normalizing the total dimming brightness $I_{\text{cu,diff}}(t_{i})$ by the area $A(t_{n})$.
These parameters are calculated from base-difference images, i.e. describing the intensity decrease with respect to the intensity level before each event. 
In addition we also define the so-called impulsive phase of the dimming, via the highest peak identified in its area growth rate profile. During this time range the dimming region reveals the strongest growth. 

Details on the dimming parameters, their distributions and a statistical comparison to the associated flare for the whole event sample, are given in paper I.

\subsection{CME kinematics}\label{sec:cme}
The time evolution of the CMEs is studied by measuring the position of the leading edge in the CME’s main propagation direction from STEREO/EUVI, COR1 and COR 2 running-difference images. We either used an adapted version of the semi-automatic algorithm developed in \cite{Bein:2011} to identify the leading edge or detect the front manually. 
Thereby, the running-difference images were contoured with an intensity level that is higher than a certain threshold defined by the mean and the standard deviation of each image: $I_{\text{crit}}>\bar{I}+a\cdot \sigma_{I}$, where $a=0.5$ for EUVI and COR1 images and $a=0.1$ for COR2 images, respectively.
 
Several factors affect the determination of the leading front of the CME. The structure itself might change rapidly over time, making it difficult to track the same feature over several solar radii. Also the combination of different instruments with different stray light levels, and different detector sensitivities within the FOV influence the appearance of the observed white-light feature. We use the following average errors in the height determination estimated in \cite{Bein:2011} for the different instruments on-board STEREO: 0.03~R$_{\odot}$ for EUVI measurements, 0.125~R$_{\odot}$ for COR1, and 0.3~R$_{\odot}$ for COR2 data, respectively.

In order to obtain robust CME estimates of the velocity and acceleration profiles, we first smooth the height-time profile. This smoothing algorithm is based on the method described in \cite{Podladchikova:2017}. It optimizes between two criteria: the closeness of the approximating curve to the data and the smoothness of the approximating curve. A detailed description of the fitting method applied to CME kinematics and the calculation of errors for the reconstructed profiles is presented in Podladchikova et al. 2018 (in preparation).

\begin{figure*}
\centering
\includegraphics[width=0.76\textwidth]{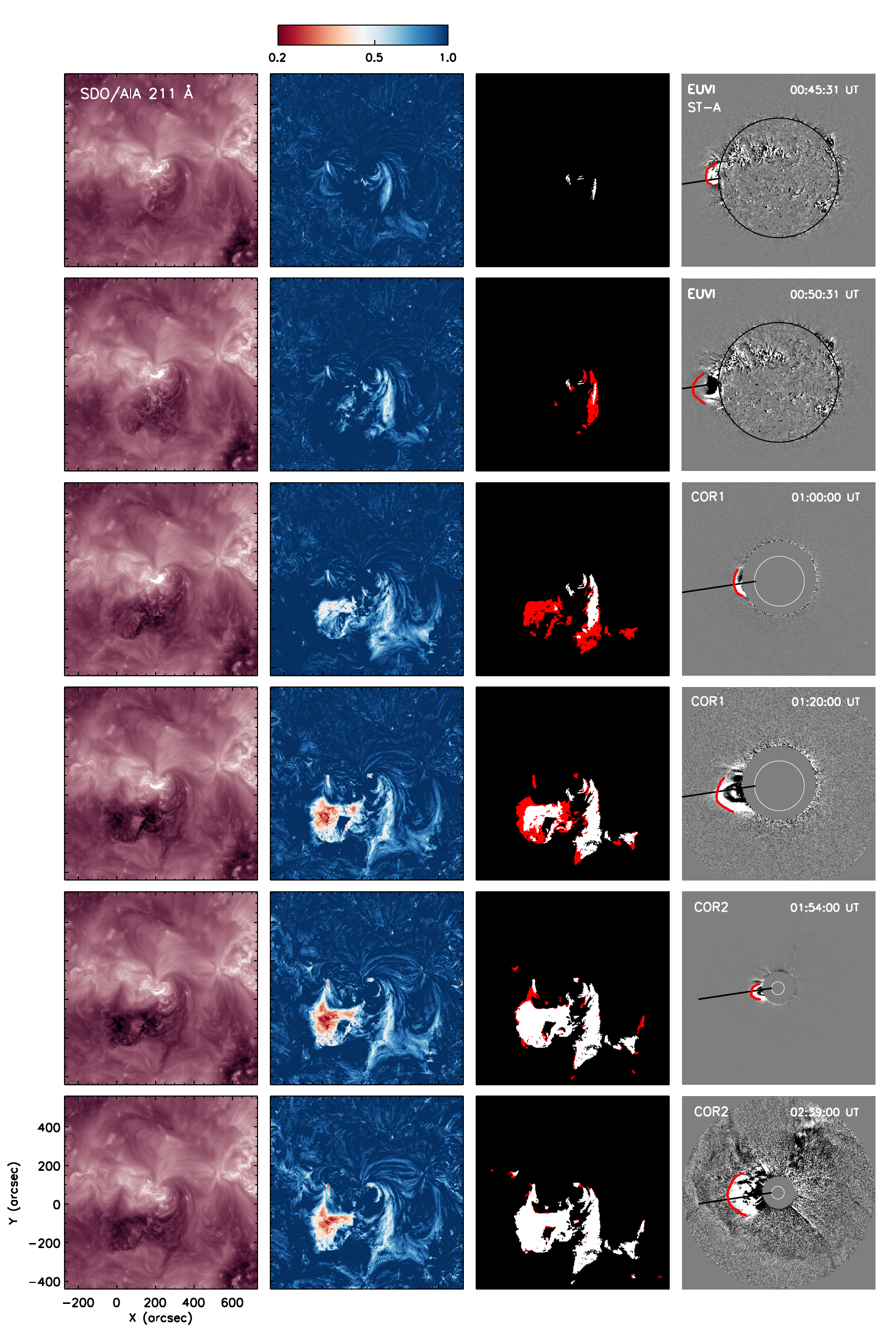}
\caption{Coronal dimming evolution and associated CME on 2011 October 2 (event no. 29). The first three columns show sequences of the coronal dimming evolution in direct SDO/211\AA~filtergrams, the corresponding logarithmic base-ratio images and the cumulative dimming masks in white (from left to right). The red pixels on top of the cumulative dimming pixel masks (third column) represent all newly detected pixels compared to the previously shown time step. The last column presents the same time steps from STEREO's point of view showing the development of the associated CME. The detected CME front is indicated in red, and the solid black line illustrates its main direction along which the height is measured (intersection between red and black line).}
\label{fig:dimming_cme_overview_20111002}
\end{figure*}

\begin{figure*}
\centering
\includegraphics[width=0.76\textwidth]{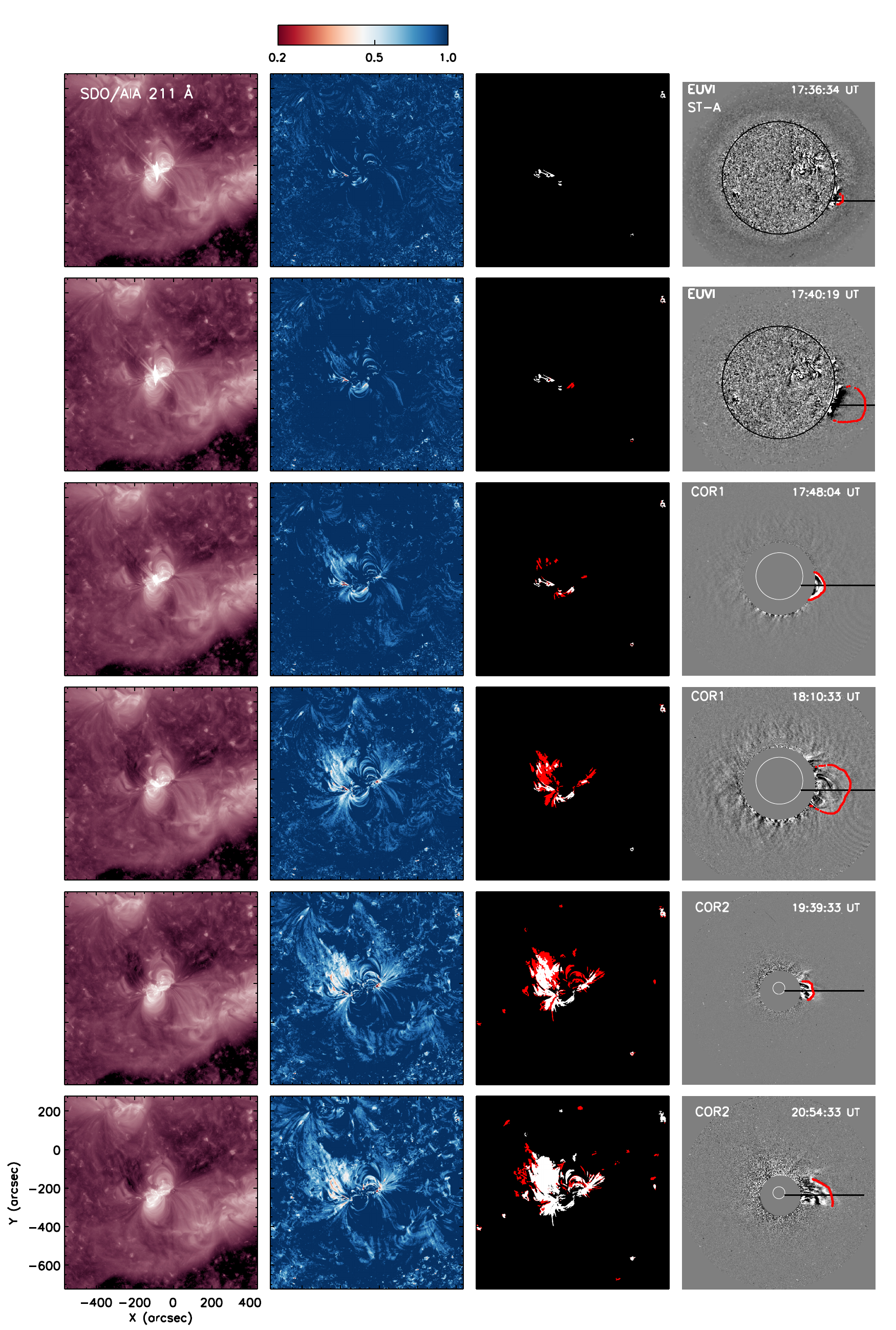}
\caption{Same as Figure~\ref{fig:dimming_cme_overview_20111002} but for event no. 6 on 2011 February 13.}
\label{fig:dimming_cme_overview_20110213}
\end{figure*}

Figures~\ref{fig:dimming_cme_overview_20111002} and \ref{fig:dimming_cme_overview_20110213} show the detection of coronal dimming regions in SDO/AIA together with the evolution of the associated CME in STEREO/EUVI, COR1 and COR2, for events no. 29 and 6. The first column shows a sequence of SDO/AIA 211\AA~images illustrating the time evolution of the dimming region. Already from the direct images, the formation of the dimming region toward the South-East can be seen. The second column presents the corresponding logarithmic base-ratio images, where regions that appear from light blue to red mark a moderate to strong intensity decrease. The third column shows all identified dimming pixels up to the given time step as cumulative dimming masks. Pixels in red represent all newly detected dimming pixels compared to the previous time step. The last column indicates the corresponding observations from STEREO/EUVI, COR 1 and COR2 of the associated CME at the simultaneous time steps. The identified leading edge of the CME is marked in red and its main propagation direction is represented by the black line. For both events, the majority of the dimming region develops in the early propagation phase of the CME, i.e. below $\sim$2.0~R$_{\odot}$.

We derive characteristic CME parameters describing the kinematics following \cite{Bein:2011}: the peak velocity $v_{\text{max}}$, the peak acceleration $a_{\text{max}}$, the acceleration duration $t_{\text{acc}}$, the height at peak velocity $h_{\text{vmax}}$, the height at peak acceleration $h_{\text{amax}}$, and the first measured height $h_{0}$.

The height $h_{0}$ at which the CME was first observed in EUVI can be interpreted as a rough estimate of the CME initiation height.
The peak velocity $v_{\text{max}}$, and the peak acceleration $a_{\text{max}}$ are defined as the maximum in the velocity- (acceleration-) time profile of the smoothed profile, respectively. The height at peak velocity $h_{\text{vmax}}$ and height at peak acceleration $h_{\text{amax}}$ were extracted from the height-time profile at the times of peak velocity and peak acceleration. The acceleration duration $t_{\text{acc}}=t_{\text{acc\_end}}-t_{\text{acc\_start}}$ is obtained using the start time $t_{\text{acc\_start}}$ and the end time $t_{\text{acc\_end}}$ of the acceleration phase, which are defined as the times when the acceleration profile drops to 10\% of its peak value. 
In addition, we also define the height of the CME when the maximum of the impulsive phase of the dimming is reached, i.e. the maximal area growth rate of the dimming, $h_{\text{dim,max}}$ and the CME height when the impulsive phase of the dimming is ending, $h_{\text{dim,end}}$. 
\subsection{CME mass}\label{sec:cme_mass}
The CME mass is calculated using the technique described in \cite{Vourlidas:2010}. We construct base-difference images, by subtracting a pre-event image that contains no other CME signature or any other disturbance. The CME and pre-event images are corrected for instrumental effects and calibrated in mean solar brightness. The obtained base-difference images should then contain only the excess brightness due to the CME under study.  
The excess brightness image is transformed to the number of electrons using the Thomson scattering equations \citep{Billings:1966} and the assumption that all electrons lie on the plane of sky. 
We calculate the CME mass by summing all pixels inside a selected region and assuming a composition of 90\% H and 10\% He. The regions are extracted either manually or defined as a sector containing the CME structure. 

\begin{figure}
\centering
\includegraphics[width=0.82\columnwidth]{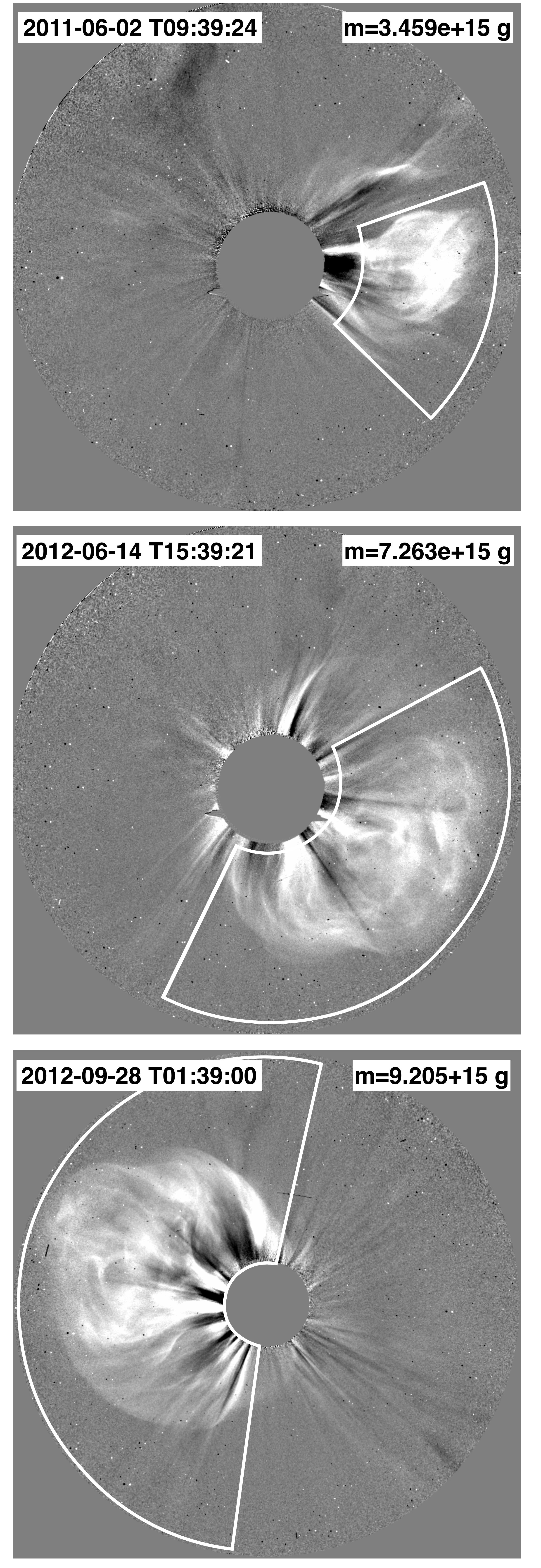}
\caption{STEREO-COR2 mass images of the CME events that occurred on 2011 June 2 (top), 2012 June 14 (middle) and 2012 September 28 (bottom). The intensity within the user-defined sector (outlined in white) is used to perform the mass calculation. The resulting value for the CME mass is given in the top right corner of each panel.}
\label{fig:mass}
\end{figure}

For our study we obtained the CME mass at one time step, when the entire CME structure is fully visible in the COR2 FOV using both methods. We calculate the mean and the standard deviations of both measurements, representing the central values, and the $1\sigma$ error bars for each event.
Figure~\ref{fig:mass} shows STEREO-COR2 mass images illustrating the calculation of the CME mass using the ``sector'' method for three example events that occurred on 2011 June 2 (top panel, no. 16), 2012 June 14 (middle panel, no. 52) and 2012 September 27 (bottom panel, no. 62). The corresponding value for the CME mass is given in the top right corner of each image.

\section{Results}
For all 62 events, we measured and analyzed the CME kinematics using either STEREO-A or \mbox{STEREO-B} data, depending from which perspective the CME could be better observed.
For 37 events a velocity profile, characterized by a clear peak in the curve, could be reconstructed from the height-time measurements. For 15 events, where the EUVI cadence was better than 5~minutes, we were able to determine reliable acceleration profiles.
The CME mass was calculated for 41 events. For the remaining 21 events, the determination of the CME region in the STEREO/COR2 observations was not possible, since the signatures of CMEs that occurred close in time are present in the images or the CMEs were too weak to be identified in the COR2 field-of-view.

Within our dataset the maximal velocities of the CMEs vary from 370--3700~km~s$^{-1}$, the maximal accelerations range from 90--3340~m~s$^{-2}$, and the CME masses have values between $2.0\times10^{14}$~g and $1.8\times10^{16}$~g, respectively. These value ranges are in agreement with former statistical studies \citep[e.g.][]{Zhang:2006,Vrsnak:2007,Bein:2011}.
Table~\ref{tab:results} lists all the CME parameters derived, together with selected dimming and flare parameters analyzed in paper I. For simplicity only the central values of each parameter are listed.
\begin{figure*}
\centering
\includegraphics[width=0.93\columnwidth]{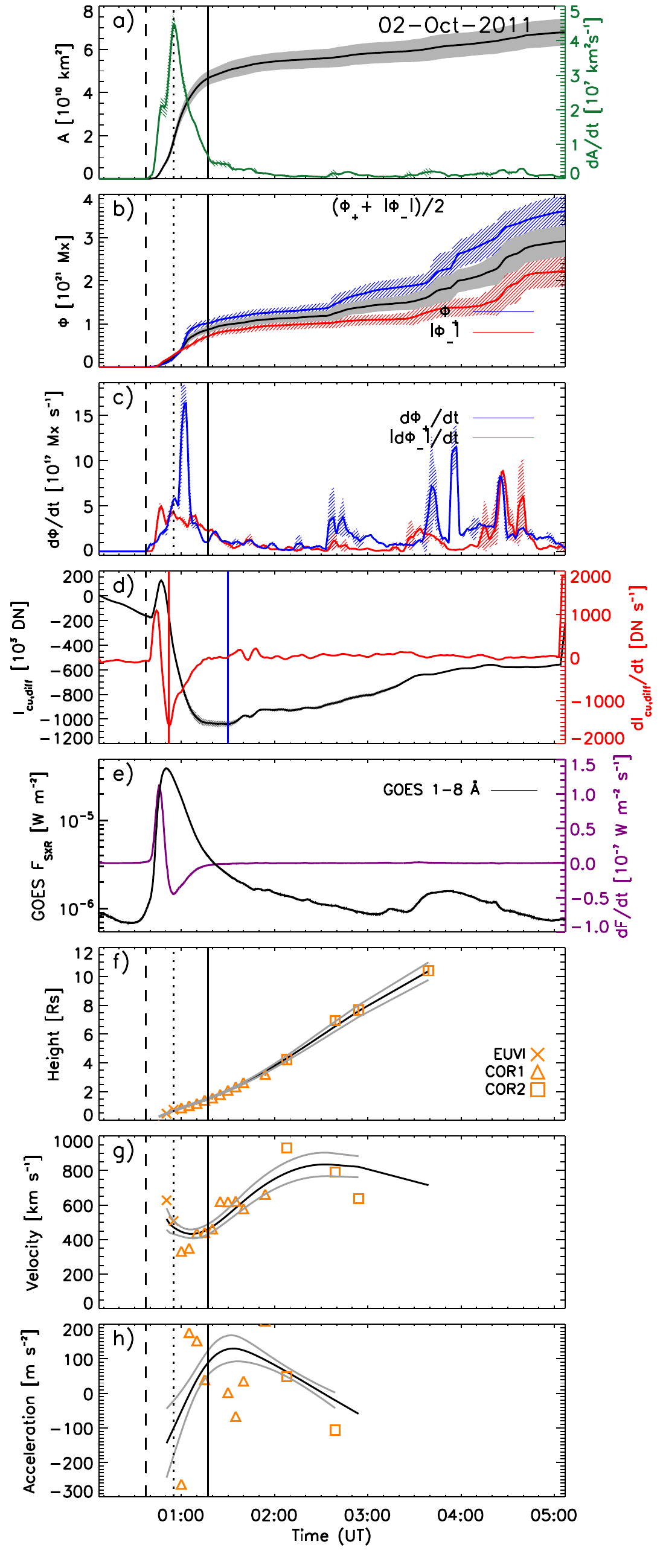}
\includegraphics[width=0.93\columnwidth]{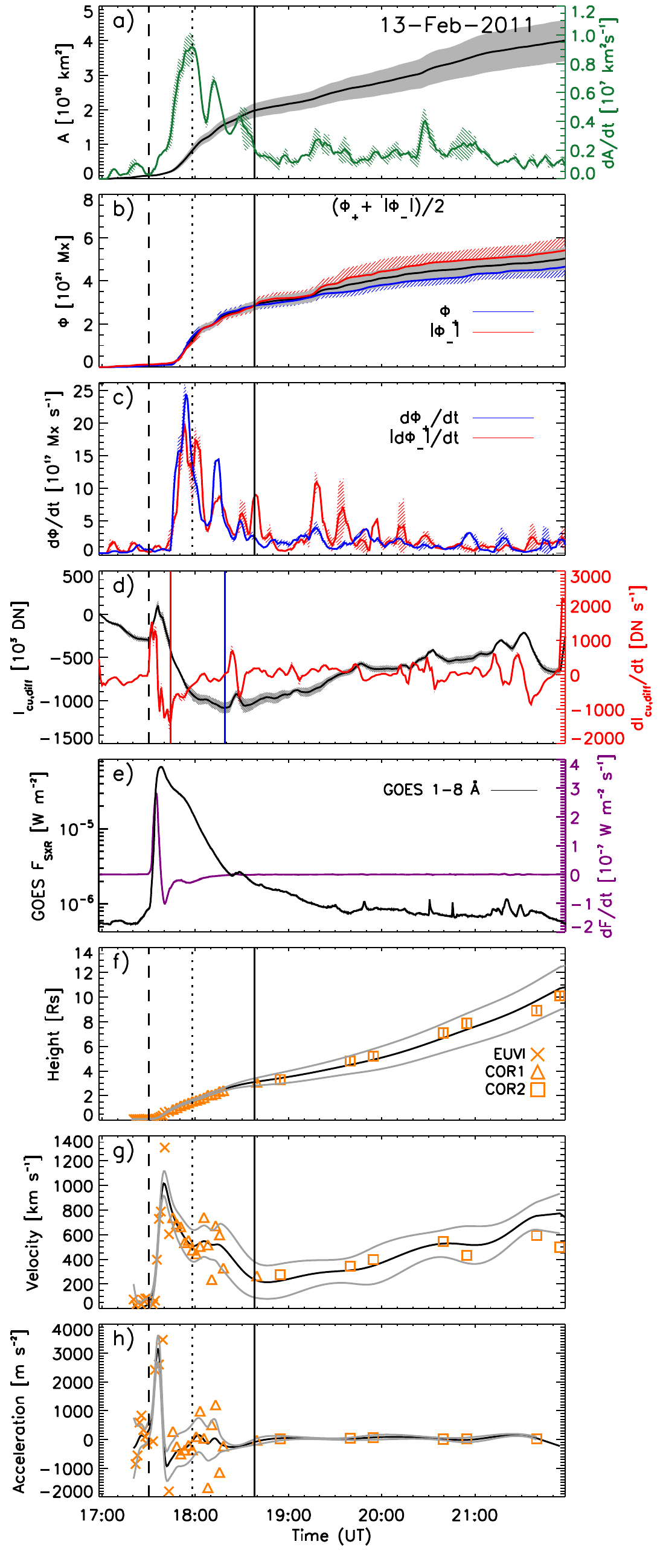}
\caption{Time evolution of dimming parameters for the dimming events that occurred on 2011 October 2 (event no. 29, left) and on 2011 February 13 (event no. 6, right): (a) the dimming area $A$ (black) and its area growth rate $\dot{A}$ (green), (b) the positive $\Phi_{+}$ (blue), the negative $\Phi_{-}$ (negative), and the total unsigned magnetic flux $\Phi$ (black), (c) the magnetic flux rates $\dot{\Phi}_{+}$ (blue) and $\dot{\Phi}_{-}$ (red), (d) the total dimming brightness $I_{\text{cu,diff}}$ (black) and its brightness change rate $\dot{I}_{\text{cu,diff}}$ (red). The shaded bands represent the $1\sigma$ error bars for each parameter. Panel (e) shows the GOES 1.0--8.0 \AA~SXR flux (black) of the associated flare and its derivative (purple) of the associated flare. The CME kinematics is shown as (f) height-time curve, (g) velocity, and (h) acceleration profile in black together with the $1\sigma$ error bars indicated by the gray lines. The original height-time measurements as well as the velocity and acceleration estimates derived from numerical differentiation are shown as orange symbols (see legend for the different symbols used for different instruments). The dashed, dotted, and solid lines mark the start time, the peak time and the end time of the impulsive phase of the dimming, respectively.}
\label{fig:20111002_profile}
\end{figure*}

Figure~\ref{fig:20111002_profile} shows, exemplary for event nos. 29 and 6, the time evolution of the selected dimming parameters (panels (a)-(d)) together with the GOES SXR flux of the associated flare and its derivative (panel (e)) and the associated height-time (panel (f)), velocity-time (panel (g)) and acceleration-time profiles (panel (h)). 
Panel (a) shows the cumulative dimming area $A$ (black) and its time derivative, the area growth rate $\dot{A}$ (green). For both events, the characteristic peak in the area growth rate occurs co-temporal with the rise in the GOES SXR flux for both events. The start time, maximum time and end time of the impulsive phase of the dimming is indicated by the dashed, dotted and solid vertical lines. Panel (b) and (c) show the positive $\Phi_{+}$, absolute negative $|\Phi_{-}|$, and total unsigned magnetic flux $\Phi$ (black) within the dimming region and the corresponding magnetic flux rates. For both events the amount of positive and negative magnetic flux is roughly balanced within the impulsive phase of the dimming.
The time evolution of the total dimming brightness $I_{\text{cu, diff}}$ and its time derivative the brightness change rate $\dot{I}_{\text{cu,diff}}$, are shown in panel (d).
The time evolution of the CME height (panel (f)), velocity (panel (g)), and acceleration (panel (h)) are also shown.

\begin{figure}
\centering
\includegraphics[width=1.0\columnwidth]{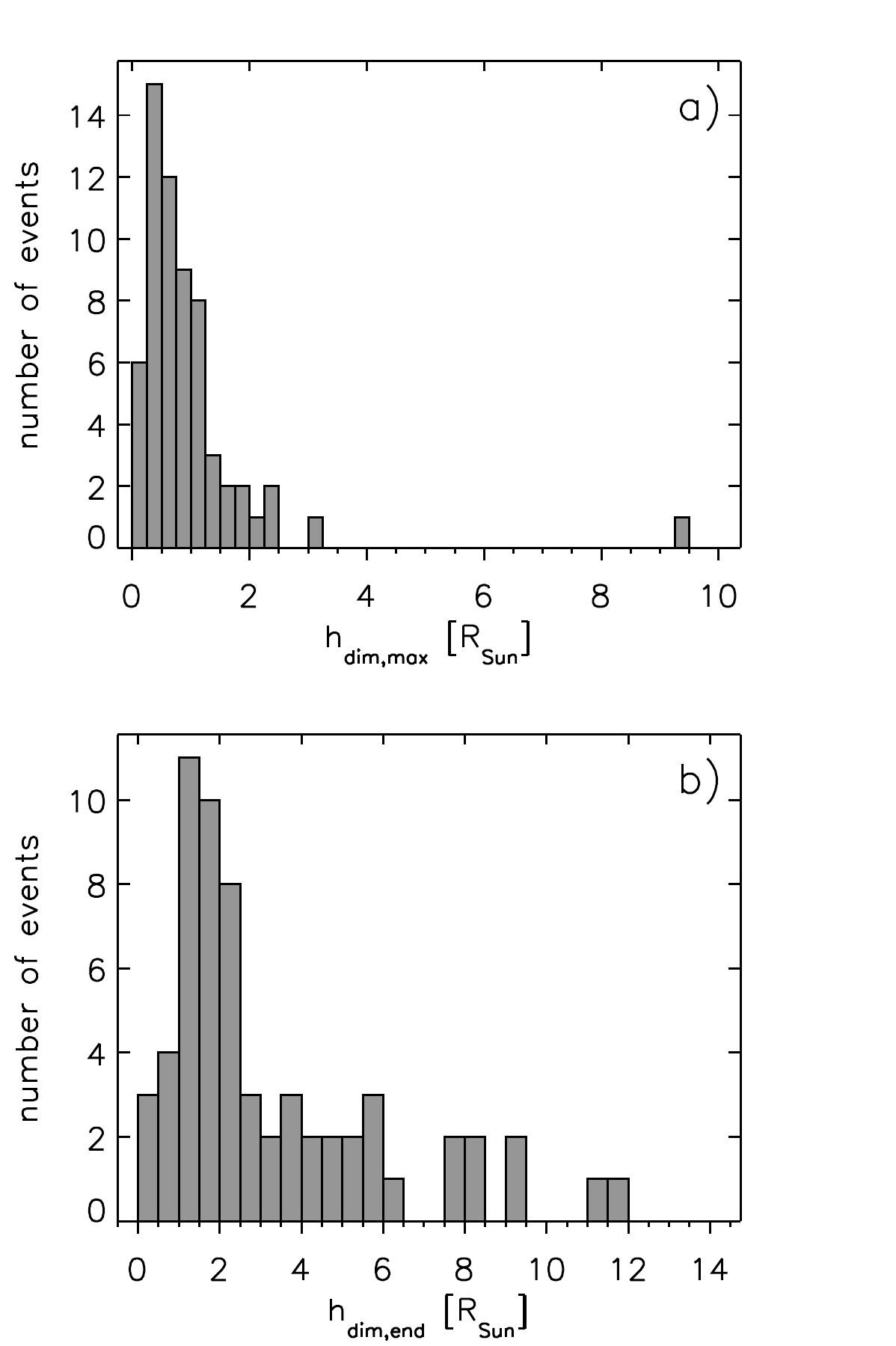}
\caption{Distributions of the height of the CME (a) at the time of the maximum and (b) at the end of the impulsive phase of the dimming.}
\label{fig:height_dimming}
\end{figure}

The velocity profile for the CME event that occurred on 2011 October 2 shows a rather slow rise and reaches its maximum speed of 800~km~s$^{-1}$ at a height of 10~R$_{\odot}$, in COR2. This behavior is in contrast to the time evolution of the velocity for the second event shown (2011 February 13). Here, the peak velocity of $\sim$ 1000~km~s$^{-1}$ is reached below 2~R$_{\odot}$ and reduces quite fastly to about 300~km~s$^{-1}$ later on.
The early acceleration phase of the CME is properly resolved for event no. 6 (2011 February 13) showing a impulsive rise to a peak acceleration of 3000~m~s$^{-2}$. The main acceleration phase of the CME also occurs within the time range of the impulsive phase of the dimming, marked by the dashed and solid vertical lines. For event no. 29 the peak acceleration of $\sim$150~m~s$^{-2}$ is reached higher up, at about 2~R$_{\odot}$ and after the end of the impulsive phase of the dimming.
 \renewcommand{\arraystretch}{0.8}
\begin{ThreePartTable}
\begin{TableNotes} \footnotesize
 \item \textbf{Note.} For each event we list the date, start time of the associated flare (from the GOES flare catalog, or derived from the GOES SXR flux using the same criteria as in the GOES flare catalog), the peak of the GOES SXR flux $F_{P}$, and the total unsigned reconnected flux $\Phi_{\text{rbn}}$ extracted from flare ribbon observations by \cite{Kazachenko:2017}. The coronal dimming is represented by its area $A$, the total unsigned magnetic flux $\Phi$, the total unsigned magnetic flux rate $\dot{\Phi}$, the mean unsigned magnetic flux density $\bar{B}_{\text{us}}$, and the total dimming brightness $I_{\text{cu,diff}}$. The associated CME is characterized by its mass $m_{\text{CME}}$, the peak velocity $v_{\text{max}}$, and the peak acceleration $a_{\text{max}}$, as well as the height of the CME at the time of the maximum of the impulsive phase of the dimming $h_{\text{dim,max}}$.
\end{TableNotes}
\begin{longtable*}{p{0.3cm}lp{0.7cm}p{1.3cm}p{0.7cm}p{1.1cm}p{0.7cm}p{0.9cm}p{0.9cm}p{0.9cm}p{0.9cm}p{1.1cm}p{0.7cm}p{0.9cm}}
 \caption{Results of characteristic dimming parameters together with basic flare and CME quantities} \label{tab:results}\\
 \toprule
 \# & Date &Start time &$F_{P}$ [W~m$^{-2}$] &$\Phi_{\text{rbn}}$ [Mx] & $A$ [km$^{2}$]& $\Phi$ [Mx]& $\dot{\Phi}$ [Mx~s$^{-1}$] &$\bar{B}_{\text{us}}$ [G] & $I_{\text{cu,diff}}$ [DN] & $m_{\text{CME}}$ [g]& $v_{\text{max}}$ [km~s$^{-1}$] & $a_{\text{max}}$ [m~s$^{-2}$] & $h_{\text{dim,max}}$ [R$_{\odot}$]\\
 & &[UT]& &($10^{21}$) &($10^{10}$) &($10^{21}$)&($10^{18}$)& &($10^{5}$)&($10^{15}$)& & & \\ \midrule

1&20100716&15:13&1.82E-07&-&1.16&0.39&0.32&32.19&-2.62&-&370.9&-&0.30 \\
2&20100801&07:37&3.24E-06&2.96&9.33&8.29&4.31&57.08&-47.98&4.05&1260.8&-&0.46 \\
3&20100807&17:54&1.04E-05&4.75&3.97&2.37&2.34&32.59&-25.25&6.70&961.5&-&0.53\\
4&20101016&19:06&3.15E-05&2.76&1.30&0.93&0.91&38.28&-8.05&-&-&-&0.54\\
5&20101111&18:52&9.26E-07&-&0.37&0.21&0.27&52.41&-&-&-&-&0.54\\
6&20110213&17:28&6.68E-05&5.12&1.99&2.90&2.23&141.92&-10.85&2.51&1015.2&3164.6&1.60\\
7&20110214&02:34&1.68E-06&0.76&0.20&0.52&1.45&278.40&-&-&-&-&0.38\\
8&20110214&04:28&8.34E-06&2.48&1.06&1.64&0.86&128.39&-&0.47&-&-&1.49\\
9&20110214&17:20&2.24E-05&-&1.99&2.97&2.09&137.59&-&-&553.4&-&0.84\\
10&20110215&01:44&2.31E-04&11.60&3.60&3.80&2.07&107.77&-&6.27&1326.5&353.5&3.17\\
11&20110215&04:27&5.30E-06&-&1.14&3.68&4.25&200.62&-&-&-&-&0.44\\
12&20110215&14:32&4.88E-06&1.64&1.16&1.09&1.08&91.60&-6.98&-&912.7&-&1.07\\
13&20110307&13:44&1.99E-05&5.18&2.94&1.45&0.77&40.13&-9.33&4.35&1104.9&825.5&0.81\\
14&20110308&18:52&1.00E-07&-&0.27&0.12&0.14&48.38&-0.59&-&-&-&1.23\\
15&20110325&23:08&1.02E-05&1.56&0.76&0.12&0.24&20.80&-1.56&-&579.9&-&0.47\\
16&20110602&07:21&3.78E-06&1.70&3.12&3.15&1.97&75.70&-15.98&3.60&1344.2&-&9.41\\
17&20110621&01:21&7.75E-06&1.13&5.24&3.15&1.25&66.12&-38.55&6.97&1000.3&-&0.47\\
18&20110703&00:00&9.54E-07&-&1.11&0.93&1.10&64.85&-3.57&0.97&1257.9&-&0.76\\
19&20110711&10:46&2.63E-06&0.26&4.19&2.05&0.75&52.10&-9.61&1.49&530.7&-&2.34\\
20&20110802&05:58&1.49E-05&7.07&2.75&2.17&1.68&75.49&-10.45&6.98&-&-&2.30\\
21&20110803&13:17&6.08E-05&7.61&4.20&4.50&3.15&74.30&-28.83&7.69&1609.2&-&0.94\\
22&20110906&01:35&5.38E-05&3.26&5.67&3.85&3.09&68.61&-25.44&4.95&929.4&-&1.23\\
23&20110906&22:12&2.14E-04&5.92&8.45&7.97&7.28&79.12&-58.41&10.85&1154.3&-&0.78\\
24&20110908&15:32&6.75E-05&7.33&1.39&2.94&2.30&113.52&-&0.22&369.7&-&1.16\\
25&20110926&14:36&2.62E-05&6.42&1.97&1.54&0.86&61.05&-18.52&-&-&-&1.60\\
26&20110927&20:43&6.44E-06&1.97&3.15&2.50&2.62&89.62&-17.06&-&-&-&0.18\\
27&20110930&03:36&7.73E-06&1.89&3.55&1.02&0.54&46.11&-&-&-&-&0.06\\
28&20111001&09:20&1.28E-05&3.60&5.43&1.33&0.76&37.34&-10.95&-&574.3&-&0.41\\
29&20111002&00:37&3.92E-05&2.42&4.67&0.87&0.97&31.90&-10.42&2.43&835.6&129.9&0.66\\
30&20111002&21:20&7.65E-06&2.63&3.27&0.90&0.46&33.36&-6.27&1.73&630.7&-&0.39\\
31&20111010&14:29&4.82E-06&0.32&0.13&0.38&0.56&195.89&-1.86&-&-&-&0.34\\
32&20111115&-&-&-&0.13&0.02&0.07&20.80&-0.38&-&-&-&0.09\\
33&20111124&23:56&1.50E-06&-&3.05&2.06&1.06&50.60&-22.38&3.02&-&-&0.43\\
34&20111213&03:07&8.14E-07&-&0.57&0.20&0.27&43.83&-1.63&-&-&-&0.58\\
35&20111222&01:56&5.48E-06&1.59&1.13&1.33&1.73&107.33&-5.53&2.37&-&-&0.84\\
36&20111225&08:49&5.57E-06&1.83&2.71&1.90&1.41&64.26&-8.97&-&-&-&0.70\\
37&20111225&18:11&4.14E-05&4.45&2.16&2.06&1.99&70.12&-16.38&5.90&-&-&1.08\\
38&20111225&20:23&8.02E-06&1.43&0.80&1.39&1.16&122.92&-8.55&-&-&-&1.81\\
39&20111226&02:13&1.52E-05&2.55&1.68&1.22&1.12&51.44&-6.33&-&707.4&193.3&0.34\\
40&20111226&11:22&5.76E-06&1.09&1.86&1.70&1.90&60.88&-12.43&4.60&1016.3&327.3&0.57\\
41&20120123&03:38&8.76E-05&17.20&4.78&2.99&2.88&40.00&-25.96&12.45&1992.9&-&0.97\\
42&20120307&00:02&5.43E-04&30.40&6.66&8.31&8.66&77.11&-&18.35&3694.8&3333.8&1.04\\
43&20120309&03:21&6.36E-05&14.50&3.30&3.42&4.12&88.70&-14.67&7.02&1250.7&-&0.22\\
44&20120310&17:15&8.49E-05&16.90&4.01&7.82&5.59&107.49&-34.09&10.83&1653.0&1932.6&1.15\\
45&20120314&15:07&2.82E-05&3.09&2.83&2.30&1.88&75.43&-23.32&3.41&-&-&0.25\\
46&20120317&20:32&1.37E-05&1.32&0.93&0.56&0.53&56.12&-4.44&0.20&-&-&0.50\\
47&20120405&20:49&1.59E-06&1.28&2.91&0.87&1.07&23.02&-12.21&5.95&-&-&0.23\\
48&20120423&17:37&2.08E-06&0.60&0.34&0.13&0.24&37.82&-1.91&1.91&872.6&-&0.02\\
49&20120511&23:02&3.24E-06&3.17&5.28&3.28&1.73&65.16&-17.40&3.37&1163.8&1655.8&0.29\\
50&20120603&17:48&3.44E-05&-&3.85&1.97&1.18&35.83&-13.60&3.68&-&208.8&1.11\\
51&20120606&19:53&2.19E-05&2.05&3.41&0.82&0.69&27.92&-10.92&3.13&770.5&-&0.63\\
52&20120614&12:51&1.92E-05&3.88&4.20&10.82&4.03&136.62&-31.12&7.76&1436.7&385.5&1.30\\
53&20120702&10:43&5.61E-05&3.84&4.58&3.00&3.16&63.68&-20.31&4.85&565.6&85.6&0.70\\
54&20120702&19:59&3.80E-05&4.78&1.54&3.18&1.83&118.84&-15.27&3.69&-&-&0.91\\
55&20120704&16:33&1.89E-05&3.56&1.28&0.87&0.98&51.14&-12.11&7.51&1112.4&650.1&0.56\\
56&20120712&16:11&1.42E-04&8.64&3.55&9.02&3.90&121.08&-&17.80&1273.4&543.3&1.33\\
57&20120813&12:33&2.88E-06&1.16&0.73&0.11&0.28&27.12&-1.10&2.01&-&-&0.68\\
58&20120814&00:23&3.50E-06&1.04&1.10&0.60&0.52&56.20&-3.98&-&-&-&2.24\\
59&20120815&03:37&8.48E-07&-&1.08&0.35&0.29&26.81&-2.42&1.03&599.3&-&0.92\\
60&20120902&01:49&2.99E-06&1.33&4.79&6.22&1.73&104.97&-41.63&2.77&996.0&-&1.91\\
61&20120925&04:24&3.61E-06&0.42&1.29&0.63&0.77&48.48&-&-&468.2&-&0.26\\
62&20120927&23:35&3.76E-06&2.33&1.62&2.50&1.42&88.48&-8.49&9.38&1495.5&863.5&0.61\\
\bottomrule
\insertTableNotes\\
\end{longtable*}
 \end{ThreePartTable}
Due to the simultaneous observations of coronal dimmings and CMEs, we are also able to investigate parameters that relate both phenomena. 
Figure~\ref{fig:height_dimming} shows the distribution of the CME height related to different stages of the dimming development. Panel (a) presents the histogram of the CME height at the peak area growth rate of the dimming $h_{\text{dim,max}}$. 
Values range from 0.02~R$_{\odot}$ to 9.4~R$_{\odot}$, with a mean of $0.7\pm0.6$~R$_{\odot}$ and a median of $1.0\pm1.3$~R$_{\odot}$. The distribution shows a well-defined peak at 0.25--0.5~R$_{\odot}$ and more than 90\% of the events reveal heights $<$2.0~R$_{\odot}$. 
Panel (b) shows the histogram of the CME height at the end of the impulsive phase of the dimming, when the main development phase of the dimming is over and its final extent is reached. $h_{\text{dim,end}}$ ranges from 0.4~R$_{\odot}$ to 12.0~R$_{\odot}$. The mean of the distribution lies at $3.3\pm2.8$~R$_{\odot}$ and the median is $2.2\pm2.0$~R$_{\odot}$, respectively. At the end of the impulsive phase of the dimming, for 70\% of the events, the propagating CME was still observed below 4.0~R$_{\odot}$ and the distribution peaks at 1.0-1.5~R$_{\odot}$. 
 
\subsection{Correlations between coronal dimming and CME parameters}
\begin{figure}
\centering
\includegraphics[width=1.0\columnwidth]{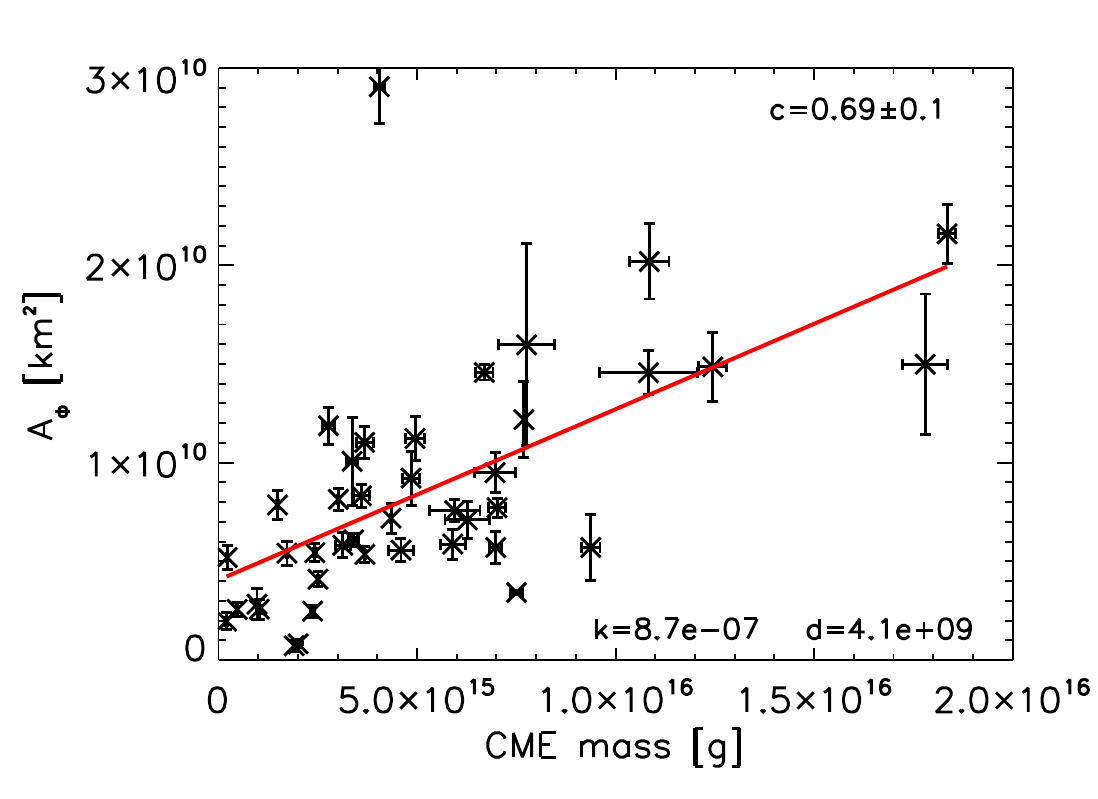}
\caption{Area of the coronal dimming $A_{\Phi}$ against the CME mass $m_{\text{CME}}$ in linear space. The red line represents the linear regression fit to all data points. The corresponding correlation coefficient is given in the top-left corner.}
\label{fig:cme_dimming1}
\end{figure}

\begin{figure}
\centering
\includegraphics[width=1.0\columnwidth]{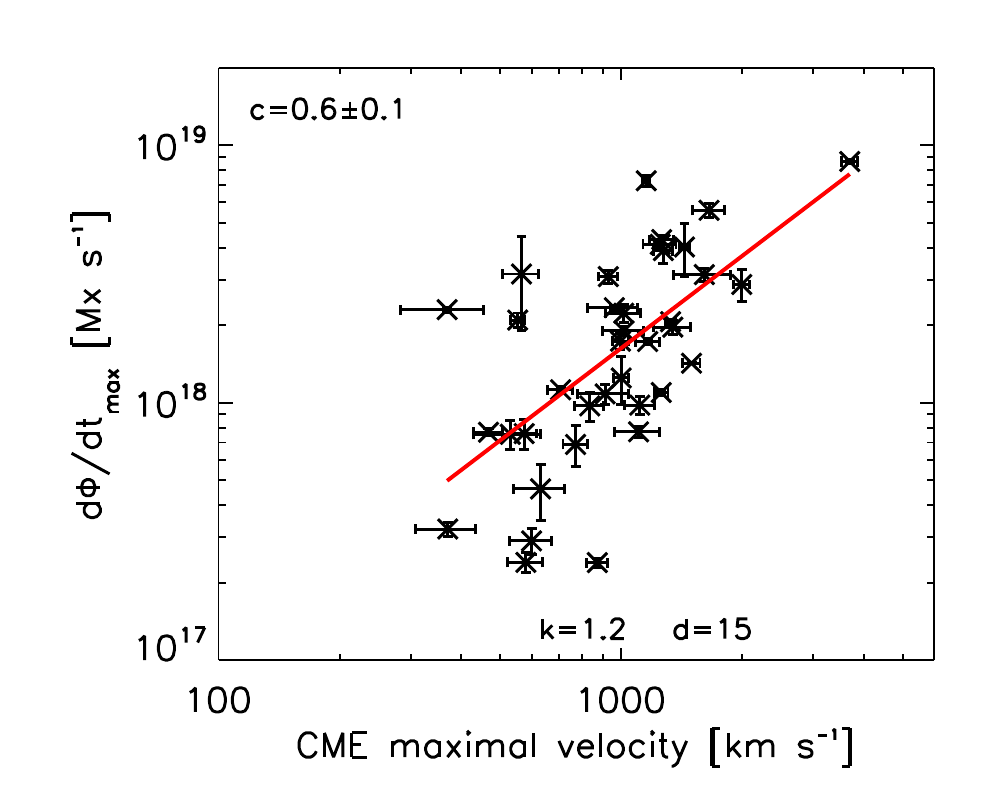}
\caption{Correlation plot between the total magnetic flux rate $\dot{\Phi}$ covered by the coronal dimming region and the maximal velocity of the CME $v_{\text{max}}$.}
\label{fig:cme_dimming2}
\end{figure}

\begin{figure}
\centering
\includegraphics[width=1.0\columnwidth]{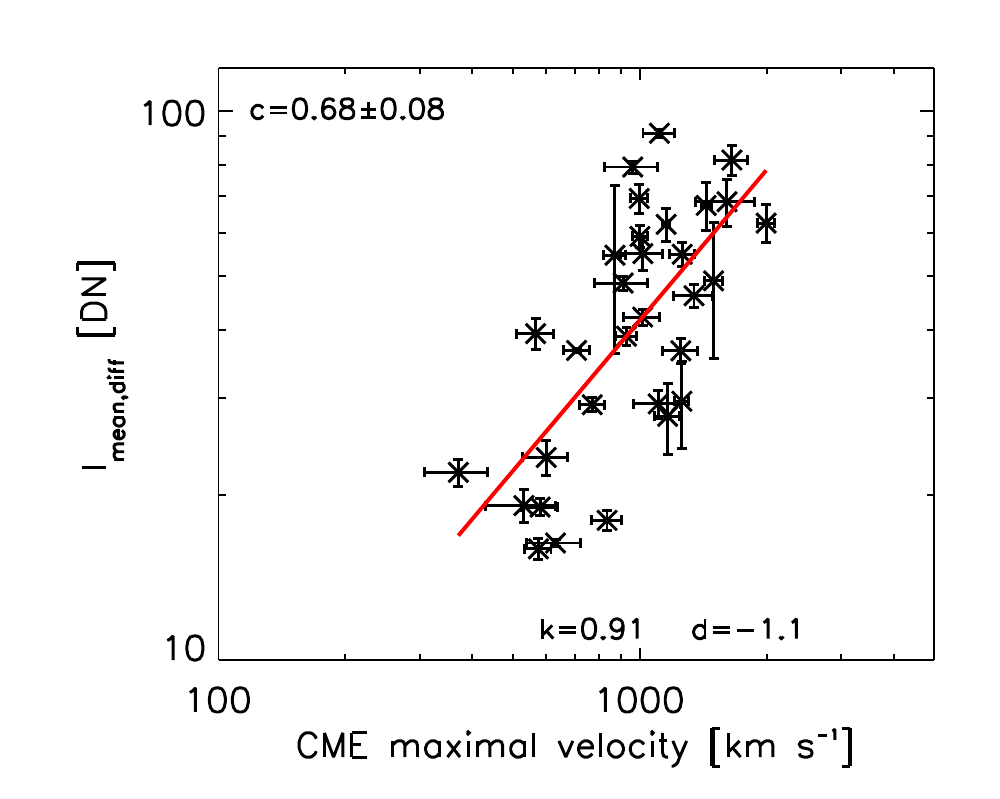}
\caption{The absolute mean intensity of the dimming $|\bar{I}_{\text{cu,diff}}|$ against the maximal velocity of the CME $v_{\text{max}}$.}
\label{fig:int_vmax}
\end{figure}

\begin{figure}
\centering
\includegraphics[width=1.0\columnwidth]{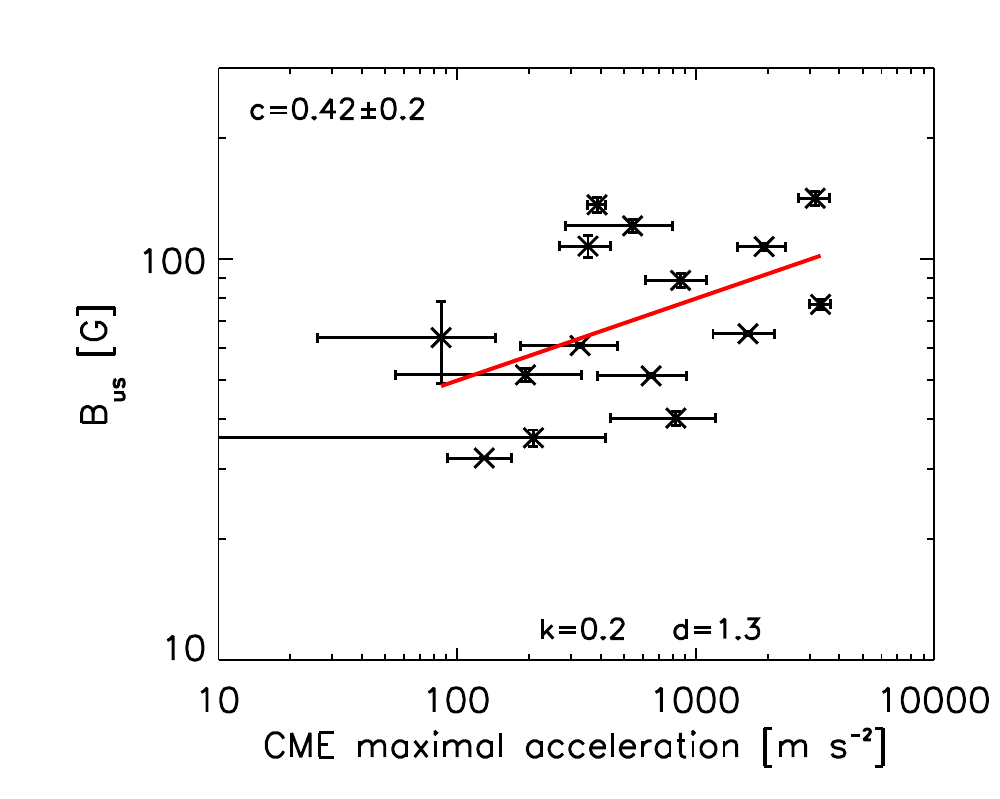}
\caption{Scatter plot of the mean unsigned magnetic field density $\bar{B}_{\text{us}}$ in the dimming region against the maximal acceleration of the CME $a_{\text{max}}$.}
\label{fig:mag_dens_amax}
\end{figure}

Figures~\ref{fig:cme_dimming1} -- \ref{fig:mag_dens_amax} show the most significant correlations we obtained between coronal dimming and CME parameters (as defined in Section~\ref{sec:method}).  
The red lines in each scatter plot represent the linear fit to the total distribution in $\log$--$\log$ space, 
\begin{equation}
\log(Y)=k\log(X)+d \;,
\end{equation}
where $k$ and $d$ represent the coefficients of the regression line, respectively.
Note, that the relationship of coronal dimming parameters with the CME mass is presented in linear space.
The corresponding correlation coefficients are given in each panel.

Figure~\ref{fig:cme_dimming1} shows the dimming area $A_{\Phi}$ against the mass of the CME $m_{\text{CME}}$. The correlation coefficient results in $c=0.69\pm0.10$, representing a strong linear relationship. The larger the area of the dimming in the low corona, the more mass the associated CME contains. The total unsigned magnetic flux $\Phi$ and the absolute total brightness of the dimming $|I_{\text{cu,diff}}|$ show a similar strong correlation with the CME mass, i.e. $c=0.67\pm0.10$ and $c=0.60\pm0.10$, respectively. These parameters were identified as first-order dimming parameters in paper I, as they reflect properties of the total dimming region at its final extent.

Figure~\ref{fig:cme_dimming2} shows the dependence of the peak of the total magnetic flux rate $\dot{\Phi}$ of the dimming and the peak velocity of the CME $v_{\text{max}}$. The scatter plot reveals a high correlation of $c=0.60\pm0.10$, indicating that the faster the total unsigned magnetic flux within the dimming region is growing per second, the higher the velocity that is reached by the CME. The area growth rate of the dimming $\dot{A}_{\Phi}$ is also strongly related to the speed ($c=0.54\pm0.10$). 
Both dimming parameters belong to the class of second-order parameters and represent the dynamics of the dimming evolution, i.e. they quantify how fast the coronal dimming is changing/growing over time (see paper I). 

Figure~\ref{fig:int_vmax} shows the absolute mean intensity of coronal dimmings $|\bar{I}_{\text{cu,diff}}|$ against the maximal speed of the associated CME $v_{\text{max}}$. The strong correlation of $c=0.68\pm0.08$ indicates that the darker the mean intensity of the dimming region the faster the associated CME. 

Although we would expect that the dimming dynamics reflects the early acceleration phase of the associated CME, no significant correlations of dimming parameters with the maximal CME acceleration could be derived. 
However, Figure~\ref{fig:mag_dens_amax} shows the absolute mean unsigned magnetic field density $\bar{B}_{\text{us}}$ against the maximal acceleration of the CME $a_{\text{max}}$. Although there is only a weak correlation ($c=0.42\pm0.20$) a trend is recognizable in the scatter plot, possibly indicating that a stronger mean magnetic flux density within the coronal dimming, may result in a higher acceleration of the associated CME. 

\begin{figure}
\centering
\includegraphics[width=1.0\columnwidth]{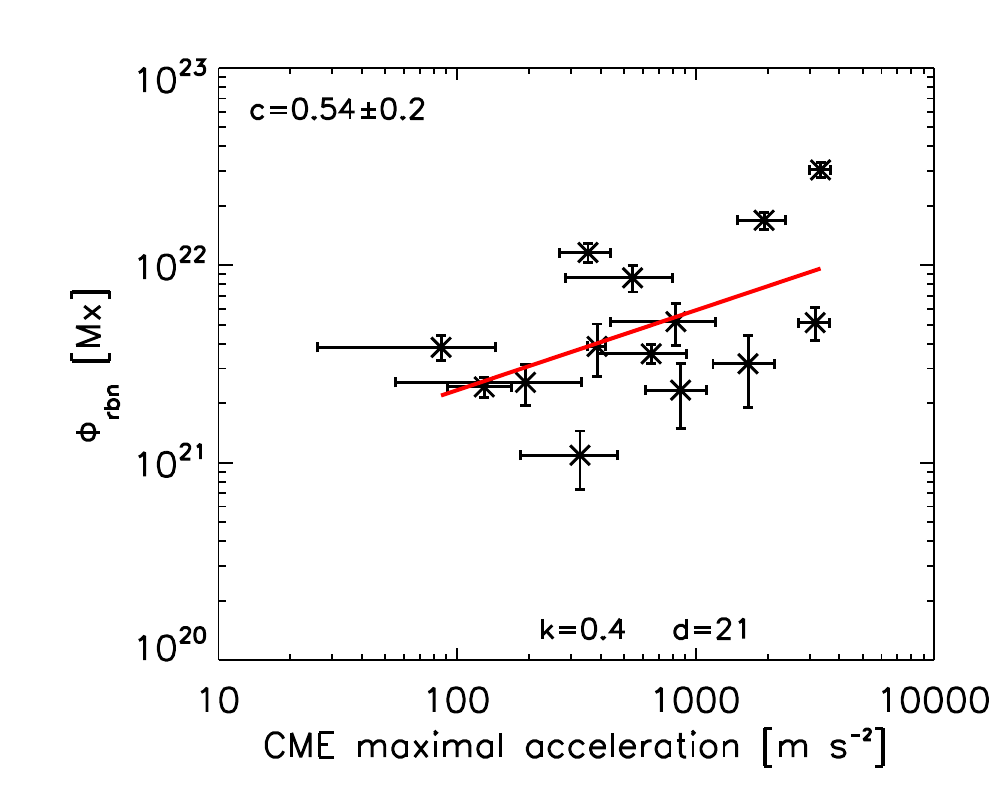}
\caption{Total reconnection flux $\Phi_{\text{rbn}}$ derived from flare ribbon observations by \cite{Kazachenko:2017} against the maximal acceleration of the CME $a_{\text{max}}$.}
\label{fig:phi_rbn_amax}
\end{figure}

The comparison between the CME acceleration and flare parameters also revealed a moderate correlation of $c=0.54\pm0.20$ between the total reconnection flux of the flare $\Phi_{\text{rbn}}$ (cf.~paper I) and $a_{\text{max}}$, as shown in Figure~\ref{fig:phi_rbn_amax}.  
The higher the flux that is reconnected during the associated flare the faster the acceleration of the corresponding CME. 

The initiation height $h_{0}$, the acceleration duration $t_{\text{acc}}$, the height at peak velocity $h_{\text{vmax}}$ and the height at peak acceleration $h_{\text{amax}}$ do not show a significant correlation with any coronal dimming parameter, in all cases $c<0.4$.
\subsection{Dimming--CME Temporal Relationship}
\begin{figure}
\centering
\includegraphics[width=1.0\columnwidth]{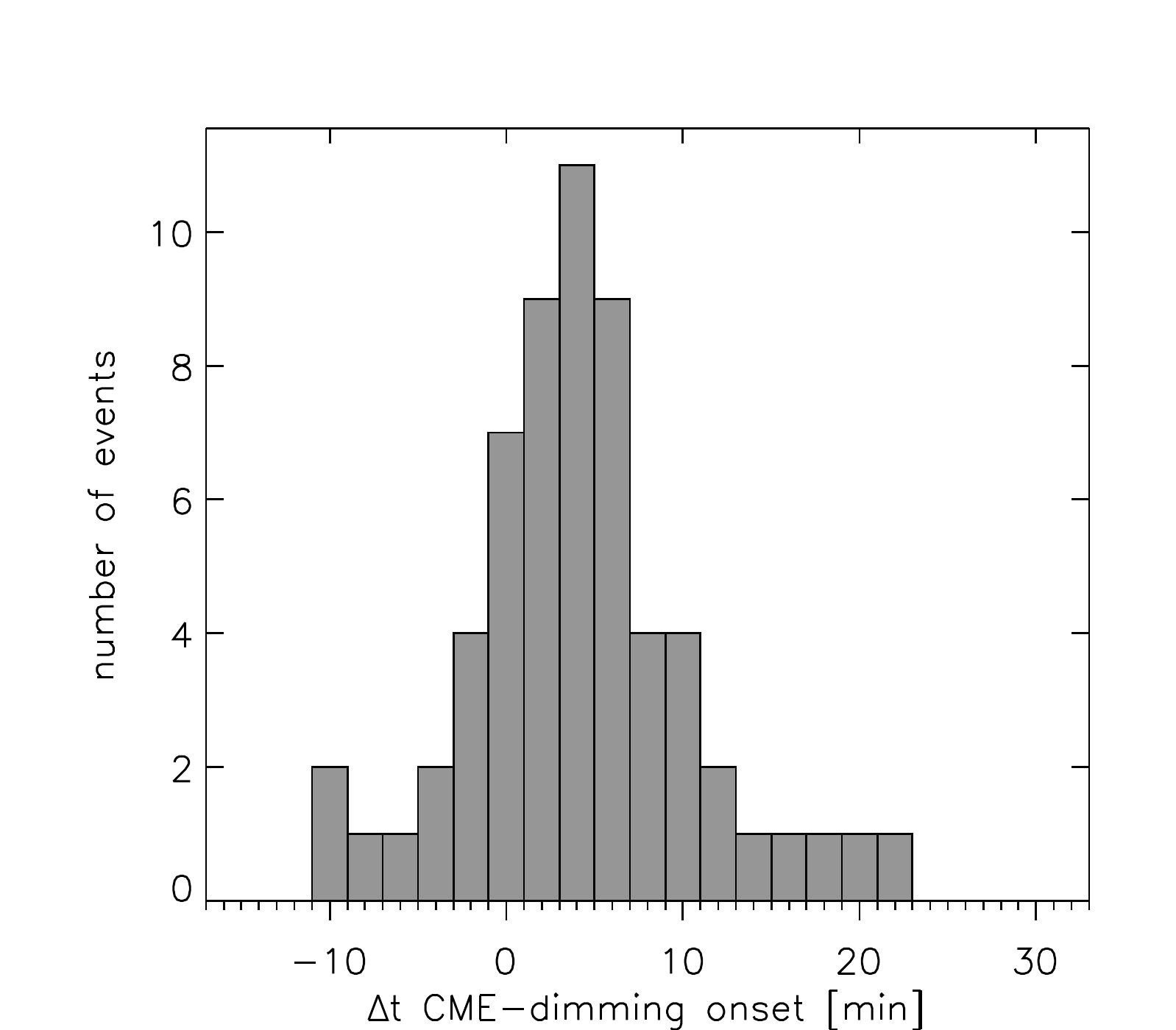}
\caption{Distribution of the time difference between the the CME onset and the start of the impulsive phase of the coronal dimming.}
\label{fig:timing}
\end{figure}

We study the relative timing between the start of the impulsive phase of the dimming and the onset of the CME rise. We define the start time of the CME as the first time a rising structure can be identified or measured in EUVI data. The dimming onset is defined as the start time of the impulsive phase of the dimming (see Section~\ref{sec:method}). Figure~\ref{fig:timing} shows the distribution of the time difference for the whole event sample. Values are ranging from \mbox{$-10.9$} to 22.2~minutes, the mean of the distribution is at $4.2\pm6.4$~minutes and the median lies at $4.2\pm4.7$~minutes, respectively.
This indicates that the initiation of the coronal dimming occurs in most of the cases before the first CME measurement in STEREO/EUVI. 
For about $55\%$ of the events the time difference lies within \mbox{$\pm$5~minutes}, which for the majority of events corresponds to the time cadence of STEREO/EUVI data. This means that within the error bars of the available measurements, the onsets of coronal dimmings and CMEs tend to occur co-temporal.

\subsection{Parameter combinations and indirect relations}
\begin{figure}
\centering
\includegraphics[width=1.0\columnwidth]{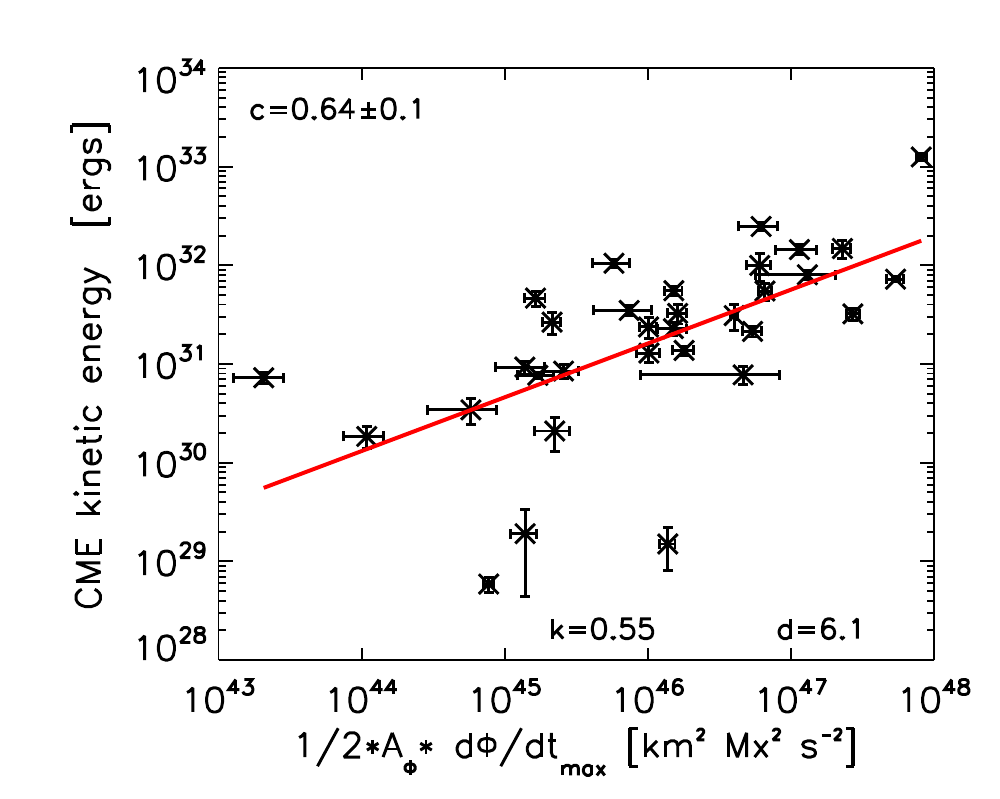}
\caption{Kinetic energy of the CME against the dimming parameter combination of the dimming area $A_{\Phi}$ and the total unsigned magnetic flux rate $\dot{\Phi}$.}
\label{fig:kinetic_energy_dimming_cme}
\end{figure}

We can also use previously identified correlations between dimming and CME parameters to provide a proxy for the kinetic energy of the CME low in the corona using first- and second-order dimming parameters.
Figure~\ref{fig:kinetic_energy_dimming_cme} shows the combination of the dimming area $A_{\Phi}$ and the total unsigned magnetic flux rate $\dot{\Phi}$ against the kinetic energy of the CMEs. We obtain a high correlation of $c=0.64\pm 0.10$, indicating that indeed a dimming parameter combination can be used to provide a proxy for the kinetic energy of the associated CME. We also note that this combination of dimming parameters correlates slightly stronger with the CME's kinetic energy than each parameter individually.

\begin{figure}
\centering
\includegraphics[width=1.0\columnwidth]{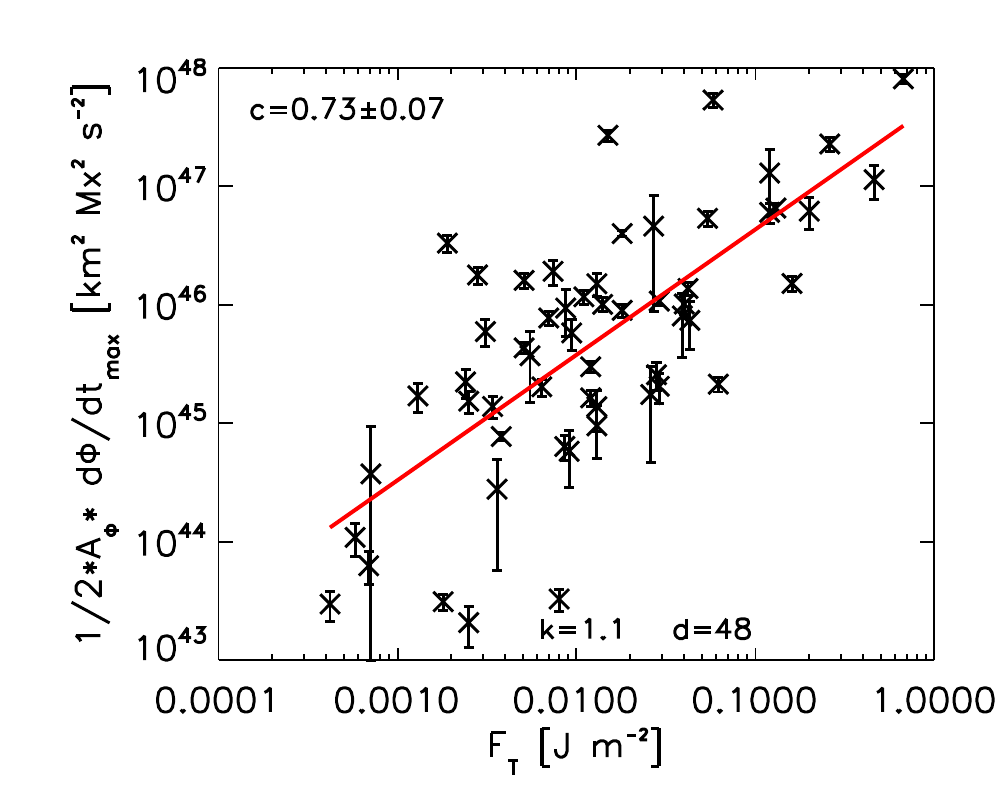}
\caption{The parameter combination of the dimming area $A_{\Phi}$ and the total unsigned magnetic flux rate $\dot{\Phi}$ against the flare fluence $F_{T}$.}
\label{fig:kinetic_energy_dimming_flare_fluence}
\end{figure}

\cite{Yashiro:2009} studied the statistical relationships between flares and CMEs observed during 1996-2007 using GOES and SOHO/LASCO data. They found the highest correlation of $c=0.62$ between the flare fluence and the kinetic energy of the CME. If coronal dimmings represent footprints of CMEs in the low corona, the kinetic energy estimated from coronal dimmings should also correlate with the flare fluence. Figure~\ref{fig:kinetic_energy_dimming_flare_fluence} reveals indeed a high correlation between the combination of first-and second-order dimming parameters and the flare fluence ($c=0.73\pm0.07$). As in Figure~\ref{fig:kinetic_energy_dimming_cme}, for the mass proxy the total dimming area $A_{\Phi}$, and for the velocity proxy, the total unsigned magnetic flux rate $\dot{\Phi}$ were used.
We also checked the correlation between the kinetic energy of the CME and the flare fluence for our event sample. The scatter plot reveals a correlation coefficient of $c=0.66\pm0.10$, similar to the value reported by \cite{Yashiro:2009}.
\subsection{Dimming Characteristics for weak CMEs}
\begin{figure}
\centering
\includegraphics[width=1.0\columnwidth]{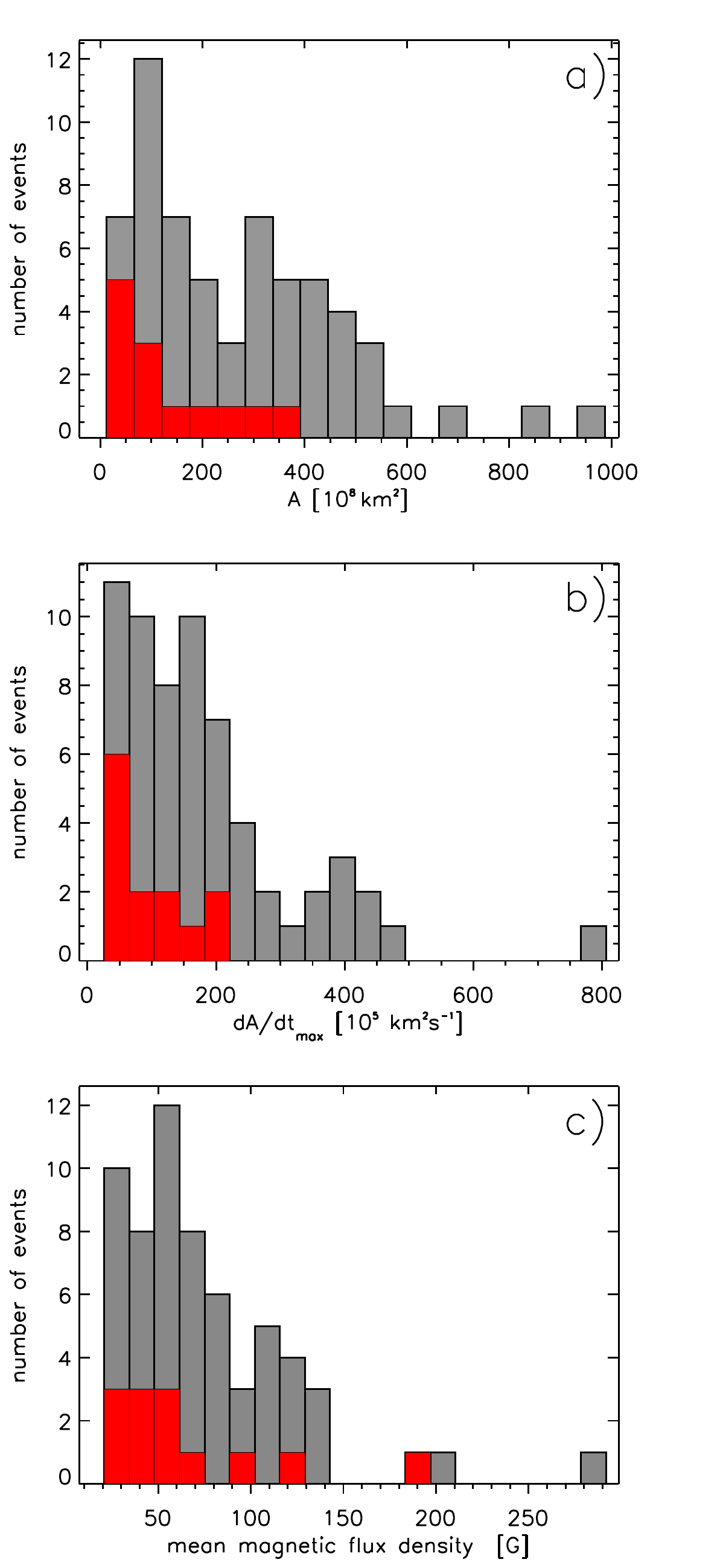}
\caption{Distribution of (a) the dimming area $A$ and (b) the maximal area growth rate $\dot{A}$  and (c) the mean unsigned magnetic flux density $\bar{B}_{\text{us}}$ for all events. The histogram of dimming events that were associated with weak CMEs is overplotted in red.}
\label{fig:histogram_weak}
\end{figure}

13 events ($\sim$20\%) within our data set rapidly dissolved over time and showed no identifiable CME signature in the outermost STEREO coronagraph, COR2. We define this subset as \textit{weak} CMEs.
Figure~\ref{fig:histogram_weak} shows the distribution of weak CMEs (red histograms) in comparison to the whole event sample (gray histograms) for three selected dimming parameters, the dimming area $A$, the maximal area growth rate $\dot{A}$, and the mean unsigned magnetic flux density $\bar{B}_{\text{us}}$. Events associated to CMEs that could not be identified in STEREO-COR2 observations are located at the left end of each distribution, i.e. are represented by the lowest values of these parameters within the sample. Also for the magnetic area, the total unsigned magnetic flux, the total dimming brightness and the corresponding time derivatives, these events group in the lower regime of parameter values. This implies that weak CMEs can already be identified low in the corona, by small, less dark coronal dimmings that include a smaller amount of magnetic flux compared to regular CMEs. 

\section{Summary and Discussion}
The statistical relationship between 62 coronal dimmings and their associated CMEs is studied using SDO/AIA and HMI, as well as STEREO/EUVI, COR1 and COR2 data.
The main findings are summarized as follows:
\begin{enumerate}
\item The majority of the dimming region develops simultaneously with the evolution of the associated CME up to an average height of $3.3\pm2.8$~R$_{\odot}$. For 90\% of the events the maximal growth rate of the dimming is reached when the CME front is still below 2.0~R$_{\odot}$ (see Figure~\ref{fig:height_dimming}). The time difference between the CME onset and the start of the impulsive phase of the dimming is $|\Delta t|<5$~minutes for 55\% and within $\pm10$~minutes for 85\% of the events (see Figure~\ref{fig:timing}), indicating a close synchronization between both phenomena (within the given observational cadence).

\item The CME mass shows the strongest correlations with first-order coronal dimming parameters, i.e. the size of the dimming region, its total unsigned magnetic flux and its total brightness \mbox{($c\sim0.6-0.7$, see Figure~\ref{fig:cme_dimming1})}.

\item The maximal speed of the CME is strongly correlated with second-order dimming parameters, i.e time derivatives of first-order parameters, such as the dimming area growth rate \mbox{($c=0.54\pm0.10$)} and the total unsigned magnetic flux rate \mbox{($c=0.60\pm0.10$, see Figure~\ref{fig:cme_dimming2})}. An even higher correlation of \mbox{$c=0.68\pm0.08$} is found with the absolute mean intensity of the dimming region (see Figure~\ref{fig:int_vmax}).

\item No significant correlation of dimming parameters with any parameter related to the CME acceleration could be derived. However, for events where high-cadence STEREO observations are available, the mean unsigned magnetic flux density of coronal dimmings $\bar{B}_{\text{us}}$ tends to be positively correlated with the maximal acceleration of the CME \mbox{($c=0.42\pm0.2$, see Figure~\ref{fig:mag_dens_amax})}. We also find that the more flux is reconnected during the associated flare, the higher is the acceleration of the associated CME ($c=0.54\pm0.2$, see Figure~\ref{fig:phi_rbn_amax}).

\item Coronal dimmings associated with weak CMEs, i.e. CMEs that show no observational signature in STEREO/COR2, are characterized by small dimming regions, that are less dark, grow slower and contain a smaller amount of magnetic flux compared to normal CMEs.
20\% of the events within our sample are represented by weak CME (see Figure~\ref{fig:histogram_weak}). 
\end{enumerate}

First-order coronal dimming parameters, reflecting properties of the total dimming region at its final extent, show the highest correlations with the CME mass. \cite{Mason:2016} also found the highest correlation with the CME mass for the dimming-related depth in the SDO/EVE lightcurves. 
Our brightness parameter $I_{\text{cu,diff}}$, extracted from spatially-resolved SDO/AIA data, reflects both the total dimming area and its intensity decrease and therefore directly corresponds to the depth in the full-irradiance SDO/EVE profile of \cite{Mason:2016}. Thus, both results are comparable and in agreement. 

\cite{Krista:2017} found a weak negative correlation ($c=-0.4$) between the peak intensity of the dimming and the CME mass.
The statistical study by \cite{Aschwanden:2017} revealed only a weak correlation of $c=0.29$ between the mass estimated from coronal dimmings and the CME mass calculated from SOHO/LASCO data.

Second-order dimming parameters, describing the dynamics of the dimming evolution and calculated as time derivatives of first-order quantities, show the strongest correlations with the maximal speed of the CME \mbox{($c\sim 0.6$)}. This is again in line with the findings of \citet{Mason:2016}, who also obtained a correlation between the intensity drop rate of SDO/EVE profiles and the CME speed.
Therefore, we conclude that the early propagation phase of CMEs is reflected by the early evolution of coronal dimmings in the low corona.
\cite{Aschwanden:2017} found only a weak correlation of $c=0.24$ between the speeds estimated from the coronal dimming and the associated CME, due to uncertainties in their maybe oversimplified EUV-dimming model.

The maximal speed of the CME shows the strongest correlation with the mean intensity of the coronal dimmings ($c=0.68\pm0.08$).
The darker the resulting dimming region is on average, the higher was the column density of the evacuated plasma during the associated CME eruption. As the plasma density in the corona decreases rapidly with height, this result indicates that faster CMEs tend to start lower in the corona.

The comparison between coronal dimming parameters and the CME acceleration revealed a moderate correlation 
between the maximum acceleration of the CME $a_{\text{max}}$ and the mean unsigned magnetic field strength $\bar{B}_{\text{us}}$ ($c=0.42\pm0.20$) in the dimming region. 
This can be interpreted that stronger fields in the CME source region are related to larger Lorentz forces, resulting in a stronger driving force accelerating the CME.

Both results, are also consistent with spectroscopic observations of Hinode/EIS shown in \cite{Jin:2009}. They found that the velocity of plasma outflows within coronal dimming regions are positively correlated with the strength of the underlying photospheric magnetic field and the relative intensity changes of coronal dimmings.

The distributions in Figure~\ref{fig:height_dimming} also show that the development of the coronal dimming in the low corona occurs co-temporal with the evolution of the associated CME up to an average height of 4.0~R$_{\odot}$. As reported in \cite{Bein:2011}, 74\% of the CMEs within their sample reached the maximum acceleration below 0.5~R$_{\odot}$. Thus, the impulsive acceleration phase of the CME should simultaneously occur with the major development phase of the coronal dimming. If higher time-cadence CME observations would be available, the dimming dynamics should therefore reflect the properties of the early CME acceleration phase.

The CME acceleration revealed the highest correlation with the total reconnected flux calculated from flare ribbon observations $\Phi_{\text{rbn}}$ ($c=0.54\pm0.2$). An even stronger relationship is found between the flare reconnection flux and the CME speed ($c=0.60\pm0.2$). This is in agreement with findings by \cite{Qiu:2005}, \cite{Deng:2017} and \cite{Tschernitz:2018} that reported a strong correlation between $\Phi_{\text{rbn}}$ and the CME speed. In a related study by \cite{Qiu:2004} also a temporal association between the total reconnected flux rate $\dot{\Phi}_{\text{rbn}}$ and the CME acceleration was found. 
It indicates, that magnetic reconnection of fields beneath the erupting structure is strongly related to to the driving of the erupting CME. In this scenario additional poloidal flux is fed into the flux rope via reconnection, thereby increasing the hoop force on the ejection leading to a greater CME acceleration and therefore also a higher maximal velocity \citep{Vrsnak:2016,Deng:2017}.

Within our data set, the average initiation height of CMEs associated with coronal dimming signatures is $0.16\pm0.13$~R$_{\odot}$ and values range from 0.01~R$_{\odot}$ to 0.84~R$_{\odot}$. This is consistent with findings in \cite{Robbrecht:2009} that modeled an initiation height of 0.4~R$_{\odot}$ above surface, for stealth CMEs, i.e. CMEs that are not associated with low coronal signatures, such as coronal dimmings.

The statistical analysis of coronal dimmings together with flares and CMEs allows us to check for possible relationships between all three phenomena.
The CME mass and the flare fluence are correlated with $c=0.62$, \citep[][present study]{Yashiro:2009}. First-order dimming parameters correlate with both, the CME mass ($c>0.6$) and the flare fluence ($c\sim0.7$, paper I).
A distinct correlation was also found between the peak velocity of the CME and the GOES SXR peak flux $F_{P}$ \citep[$c>0.5$,][present study $c=0.41\pm0.20$]{Vrsnak:2005,Maricic:2007,Bein:2012}. Second-order dimming parameters correlate again with both, the maximal speed of the CME ($c\sim0.7$) and the GOES peak flux ($c=0.6-0.7$, paper I).

If CMEs occur together with flares, they belong to the same magnetically driven event \citep[][]{Harrison:1995, Zhang:2001, Priest:2002, Webb:2012, Green:2018}. Coronal dimmings may connect these two phenomena as they statistically reflect the properties of both. On the one hand, they represent the properties of the erupting CME, i.e. its speed and mass. On the other hand the balance between the positive and negative magnetic fluxes within the dimming region ($c=0.83\pm0.04$, paper I) and the strong correlation between the flare reconnection fluxes and secondary dimming fluxes ($c=0.62\pm0.08$, paper I) indicate that roughly the same amount of magnetic flux is added to the erupting structure that is reconnected during the associated flare.
This is in agreement with the unified model proposed by \cite{Lin:2004}, where the same amount of magnetic flux leaves both ends of the reconnection site. The downward component reaches the chromosphere and creates flare ribbons at the footpoints of newly reconnected loops. The upward component is added as poloidal flux to the erupting flux rope. 

Results presented in this paper also confirm the feedback relationship between flares and CMEs \citep[e.g.][]{Vrsnak:2008,Vrsnak:2016}. It is based on observations of the close temporal relationship between the CME acceleration phase and the flare energy release \citep[e.g.][]{Temmer:2010, Berkebile:2012} as well as the flare reconnection rate \citep[e.g.][]{Qiu:2004}, and the strong correlation between the CME velocity and the total reconnection fluxes of the associated flares \citep{Qiu:2005,Deng:2017,Tschernitz:2018}.
Magnetic reconnection affects the dynamics of CMEs by the reduction of the tension of the overlying magnetic fields, the increase of magnetic pressure below the flux rope, as well as the supply of additional poloidal flux to the flux rope, prolonging the driving Lorentz forces \citep{Vrsnak:2016}.
\pagebreak

\section{Conclusions}
Coronal dimmings represent the footprints of CMEs in the low corona and provide essential additional information on their early evolution phase. We identified first-order dimming parameters, i.e. the size of the dimming, its total unsigned magnetic flux and its total brightness to be proxies for the CME mass ($c\sim0.6-0.7$). Secondary-dimming parameters, i.e. representing the time derivatives of first-order parameters, such as the area growth rate and the magnetic flux rate serve as proxies for the peak velocity of the CME ($c\sim0.6$). The maximum acceleration of the CME tends to be positively correlated with the mean unsigned magnetic field strength, consistent with a stronger Lorentz force accelerating the CME. 
 
Specific coronal dimming parameters, such as the dimming area or the total magnetic flux rate correlate with both, CME and flare quantities providing further evidence for the flare-CME feedback relationship. 
We therefore conclude that if CMEs occur together with flares, coronal dimmings statistically reflect the properties of both phenomena.
These results are of great use for estimating the impact power of Earth-directed CME events without having coronagraphic data.
\section{Acknowledgements}
K.D., A.M.V., and M.T. acknowledge funding by the Austrian Space Applications Programme of the Austrian Research Promotion Agency FFG (ASAP-11 4900217, BMVIT) and the Austrian Science Fund FWF (P24092-N16). SDO data is courtesy of NASA/SDO and the AIA, and HMI science teams. The STEREO/SECCHI data are produced by an international consortium of the Naval Research Laboratory (USA), Lockheed Martin Solar and Astrophysics Lab (USA), NASA Goddard Space Flight Center (USA), Rutherford Appleton Laboratory (UK), University of Birmingham (UK), Max-Planck-Institut f\"{u}r Sonnenforschung (Germany), Centre Spatiale de Li\`{e}ge (Belgium), Institut d'Optique Th\'{e}orique et Appliqu\'{e}e (France), and Institut d' Astrophysique Spatiale (France). 

\bibliographystyle{apj}

\end{document}